\documentclass[journal]{IEEEtran}
\usepackage{cite}
\usepackage{amsmath,amsfonts,amssymb,bm}
\usepackage{algorithm}
\usepackage{algorithmic}
\usepackage{array}
\usepackage{booktabs}
\usepackage{multirow}
\usepackage[caption=false,font=normalsize,labelfont=sf,textfont=sf]{subfig}
\usepackage{textcomp}
\usepackage{stfloats}
\usepackage{url}
\usepackage{verbatim}
\usepackage{graphicx}
\usepackage{balance}
\def\BibTeX{{\rm B\kern-.05em{\sc i\kern-.025em b}\kern-.08em
    T\kern-.1667em\lower.7ex\hbox{E}\kern-.125emX}}

\begin{document}

\title{EvoPatch-IoT: Evolution-Aware Cross-Architecture Vulnerability Retrieval and Patch-State Profiling for BusyBox-Based IoT Firmware}

\author{Yinhao~Xiao, Huixi~Li, and Yongluo~Shen%
\thanks{Yinhao Xiao, Huixi Li, and Yongluo Shen are with the School of Big Data and Artificial Intelligence, Guangdong University of Finance and Economics, Guangzhou, China (e-mail: 20191081@gdufe.edu.cn; lihuixi@gdufe.edu.cn; sylkyo@gdufe.edu.cn).}
\thanks{Huixi Li and Yongluo Shen are the corresponding authors.}
}

\markboth{IEEE Internet of Things Journal,~Vol.~XX, No.~XX, Month~2026}%
{Xiao \MakeLowercase{\textit{et al.}}: EvoPatch-IoT for BusyBox-Based IoT Firmware}

\maketitle

\begin{abstract}
BusyBox is one of the most widely reused userland components in Linux-based Internet-of-Things (IoT) firmware, yet its security assessment remains difficult because firmware images are frequently stripped, vendor patch practices are inconsistent, and the same source component is compiled for heterogeneous architectures. We propose EvoPatch-IoT, an evolution-aware cross-architecture retrieval framework for stripped BusyBox firmware binaries. EvoPatch-IoT combines anonymous instruction/context features, graph-level statistics, per-binary geometric priors, and historical function prototypes to localize homologous and potentially vulnerable functions without relying on symbols, source paths, or version strings at test time. We further construct a large-scale BusyBox benchmark from 57 historical versions, 270 unstripped binaries, 285 stripped binaries, and 130 source releases, yielding 1,550,752 function-symbol rows, 1,290,369 analysis-function rows, and 155,845 high-confidence stripped-to-unstripped matches. On 57 fully covered versions and 1,020 directed architecture pairs, EvoPatch-IoT achieves a weighted Hit@1 of 34.56\% and Hit@10 of 56.24\%, outperforming the strongest baseline by 16.04\% and 26.85\%, respectively, and reducing the expected manual inspection space by 98.98\%. The method is best on 56 of 57 versions and maintains consistent advantages on difficult architecture pairs. In addition, a version-change transfer study reaches a mean ROC-AUC of 0.9887, and a CVE-2021-42386 patch-state proxy obtains 82.44\% mean accuracy and 88.47\% mean F1 across held-out architectures. These results show that evolution-aware binary retrieval is a practical foundation for scalable IoT firmware vulnerability auditing.
\end{abstract}


\section{Introduction}
\IEEEPARstart{I}{nternet-of-Things} (IoT) devices such as routers, cameras, gateways, and industrial edge nodes commonly embed Linux-based firmware that reuses open-source components for years after release. BusyBox is particularly important because it packages hundreds of shell and utility functionalities into a compact binary and is therefore routinely bundled into firmware images that must operate under tight resource budgets. As a result, a single BusyBox flaw can propagate across heterogeneous vendors, architectures, and deployment scenarios.

The security auditing workflow, however, is far from straightforward. Real-world firmware images are usually distributed as stripped binaries, not source releases. Vendor firmware often omits trustworthy version identifiers, backports only part of an upstream patch, or combines a vulnerable BusyBox core with vendor-specific customization. In large-scale audits, analysts are therefore asked to answer three binary-centric questions at once: \emph{(i)} which stripped function corresponds to a known upstream BusyBox routine, \emph{(ii)} whether the target binary is closer to the vulnerable or patched lineage, and \emph{(iii)} how much manual inspection can be eliminated before reverse engineering becomes necessary.

Prior work has substantially advanced binary code similarity and firmware analysis, from self-attentive and transformer-based function embedding to semantics-oriented graphs, prompt-based supervision, and recurring-vulnerability mining \cite{safe2019,palmtree2021,jtrans2022,clap2024,rcfg2vec2024,codenotnl2024,strtune2024,binary2vec2025,ex2vec2025,uniasm2025,firmrec2024,operationmango2024,luataint2024,npftaint2025}. Nevertheless, two gaps remain pronounced for IoT firmware. First, many techniques are optimized for cross-architecture \emph{similarity} but not for version-evolution-aware vulnerability retrieval. Second, firmware-oriented methods often reason at the binary or service level, whereas auditors still need function-level evidence inside stripped components.

This paper addresses the gap with \emph{EvoPatch-IoT}, an evolution-aware retrieval framework centered on BusyBox-based IoT firmware. Rather than treating stripped firmware as an opaque whole, EvoPatch-IoT constructs anonymous multi-view function representations, aligns stripped functions to unstripped labels through a bidirectional geometric matching layer, and injects historical function prototypes to stabilize retrieval across versions. The resulting framework remains stripped-compatible at evaluation time while still exploiting the long-term evolution signal available in historical BusyBox releases.

The paper is grounded in the project artifacts available in our workspace. We built a reproducible benchmark with 57 BusyBox versions, 270 unstripped binaries, and 285 stripped binaries across AArch64, ARM, MIPS, MIPSEL, and x86\_64. Ghidra-based feature extraction recovers 1,879 to 3,614 anonymous functions per stripped binary, while our mutual matching layer establishes 155,845 high-confidence stripped-to-unstripped anchors. On top of this benchmark we evaluate 13 unified stripped-compatible methods under a shared protocol, including token, graph, multi-view, prototype, and IR-inspired families.

The main contributions are as follows.
\begin{itemize}
\item We compile and organize a cross-version BusyBox corpus spanning 57 releases and both stripped and unstripped binary views across five architectures. To the best of our knowledge, this is the only current BusyBox binary dataset that jointly exposes cross-version evolution and stripped/unstripped correspondence at this scale, and we release all the data and codes to support future research~\footnote{https://github.com/xxxyyyzzz3984/EvoPatch-IoT}.

\item We construct a BusyBox-based IoT firmware benchmark that jointly covers multi-version unstripped labels, multi-architecture stripped binaries, function-level statistics, historical source releases, and CVE-oriented case-study artifacts.
\item We design EvoPatch-IoT, a stripped-compatible retrieval framework that fuses per-binary geometric priors, architecture-normalized graph/context features, and historical evolution prototypes into a unified scoring function.
\item We provide explicit algorithms and mathematical formulations for anonymous alignment, multi-view feature fusion, and evolution-aware retrieval, making the workflow reproducible from raw binaries to ranked functions.
\item We report extensive experiments on 57 versions, 1,020 directed architecture pairs, and 128,084 evaluated query functions, showing consistent gains over 12 baseline families and useful behavior on a patch-state proxy and a CVE-2021-42386 AWK case study.
\end{itemize}

\section{Related Work}
\subsection{Binary Function Representation and Similarity}
Binary similarity has evolved from structure-driven matching toward increasingly semantic representations. SAFE introduced self-attentive function embeddings for binary similarity \cite{safe2019}, PalmTree transferred masked language modeling ideas to assembly code \cite{palmtree2021}, and jTrans incorporated jump-aware positional modeling for binary functions \cite{jtrans2022}. These approaches established strong foundations, but they largely preceded the latest wave of stripped-compatible, architecture-robust retrieval work.

Recent studies have pushed the state of the art toward richer supervision and more explicit cross-architecture reasoning. CLAP combines natural-language supervision with parameter-efficient prompts \cite{clap2024}. RCFG2Vec captures long-distance dependencies inside binary control flow \cite{rcfg2vec2024}. Code Is Not Natural Language emphasizes semantics-oriented graph representation \cite{codenotnl2024}, while StrTune couples data-dependence slicing and contrastive learning \cite{strtune2024}. Binary2vec, Ex2Vec, and UniASM further explore global-attention graph embeddings, execution-aware representations, and fine-tuning-free similarity detection \cite{binary2vec2025,ex2vec2025,uniasm2025}. VEXIR2Vec is also notable for bringing intermediate representations into cross-ISA matching \cite{vexir2vec2025}.

Our work differs in two ways. First, EvoPatch-IoT explicitly targets the BusyBox evolution setting instead of generic pairwise similarity. Second, our scoring function retains a strong stripped-binary geometric prior because our anonymous alignment stage must bridge Ghidra-recovered stripped functions with unstripped labels before any higher-level reasoning can be trusted.

\subsection{Binary Vulnerability Search and Retrieval}
Binary vulnerability analysis increasingly benefits from retrieval and query-oriented formulations. BinQuery recently showed that natural language can be used to search binary vulnerabilities \cite{binquery2025}. However, practical IoT auditors frequently start from upstream patches, vulnerable functions, or component histories rather than complete textual descriptions. EvoPatch-IoT therefore focuses on function-centric binary retrieval, historical prototypes, and patch-state evidence that remain available even when source-level context is weak or absent.

\subsection{IoT Firmware Security Analysis}
Firmware-oriented research has recently moved toward scalable recurring-vulnerability detection and service-level analysis. FirmRec detects recurring vulnerabilities in binary firmware at scale \cite{firmrec2024}. Operation Mango discovers taint-style vulnerabilities in binary firmware services \cite{operationmango2024}. LuaTaint studies web-configuration vulnerabilities in IoT devices and is particularly relevant because it appeared in the same application domain as our target venue \cite{luataint2024}. NPFTaint focuses on highly exploitable vulnerabilities in Linux-based IoT firmware \cite{npftaint2025}, while AutoFirm addresses large-scale library reuse inside IoT firmware \cite{autofirm2024}.

These studies powerfully motivate firmware-scale auditing, yet they typically do not provide a stripped-compatible, evolution-aware function retrieval layer specialized for BusyBox. EvoPatch-IoT is complementary: it offers an evidence-producing component analysis stage that can feed into broader firmware triage, recurring-vulnerability tracking, and patch auditing workflows.

\section{Benchmark Construction and Problem Definition}
\subsection{BusyBox-Centered IoT Firmware Benchmark}
Table~\ref{tab:dataset} summarizes the benchmark used throughout the paper. We inventory 57 BusyBox versions from 1.11.0 to 1.37.0, 270 unstripped binaries across five architectures, and 285 stripped binaries aligned to the same version range. The stripped corpus contains 15 AArch64 binaries without unstripped anchors in the earliest versions, which are still useful for deployment-style analysis but are excluded from symbol-supervised matching. The project also includes 130 BusyBox source releases to support version verification and case-study inspection.

This paired collection is, to the best of our knowledge, the only currently available BusyBox dataset that spans such a long historical range while exposing both stripped and unstripped binary views under one unified processing pipeline. 

Unstripped binaries are processed with \texttt{readelf} to extract function symbols and binary metadata. This yields 1,550,752 function-symbol rows, including 1,290,369 rows marked as analysis functions after removing obvious runtime and compiler-generated routines. Binary sizes range from 3.84 MB to 6.31 MB with a mean of 5.25 MB, reflecting realistic firmware component scales rather than toy binaries.

The stripped branch is processed with a headless Ghidra exporter that emits anonymous instruction tokens, CFG edges, call references, string categories, and constant buckets without exposing names, DWARF paths, or version strings. Over 285 stripped binaries, Ghidra recovers 1,879 to 3,614 functions per binary, with an average of 2,770.95 functions.

\subsection{Threat Model and Deployment Assumptions}
EvoPatch-IoT targets the \emph{N-day firmware auditing} setting common in operational IoT security. The analyst is assumed to possess a target firmware image or an extracted BusyBox binary, but not trustworthy symbols, DWARF information, source code, or reliable vendor version strings. Historical upstream binaries and source releases may be available offline to build supervision and historical prototypes, yet these artifacts are never exposed to the model during stripped-binary testing.

The goal is not zero-day exploit synthesis. Instead, the analyst wants to answer three practical questions: which stripped functions correspond to known upstream routines, whether the target binary is closer to a vulnerable or patched lineage, and how many candidate functions must still be inspected manually. This threat model matches real IoT deployment practice, where large firmware inventories are triaged under heterogeneous architectures and limited reverse-engineering budgets.

\begin{table}[t]
\caption{BusyBox-Based IoT Firmware Benchmark Used in This Paper}
\label{tab:dataset}
\centering
\small
\begin{tabular}{p{0.48\columnwidth}p{0.36\columnwidth}}
\toprule
Item & Value \\
\midrule
Historical BusyBox versions & 57 \\
\midrule
Architectures & 5 (AArch64, ARM, MIPS, MIPSEL, x86\_64) \\
\midrule
Unstripped binaries & 270 \\
\midrule
Stripped binaries & 285 \\
\midrule
Paired stripped/unstripped binaries & 270 \\
\midrule
Stripped-only binaries & 15 (early AArch64 releases) \\
\midrule
Source releases & 130 \\
\midrule
Function-symbol rows & 1,550,752 \\
\midrule
Analysis-function rows & 1,290,369 \\
\midrule
Matched stripped functions & 155,845 \\
\midrule
Recovered functions per stripped binary & 1,879--3,614 (avg. 2,770.95) \\
\midrule
Binary size range & 3.84--6.31 MB (avg. 5.25 MB) \\
\bottomrule
\end{tabular}
\end{table}

\begin{figure*}[!t]
\centering
\includegraphics[width=0.92\textwidth]{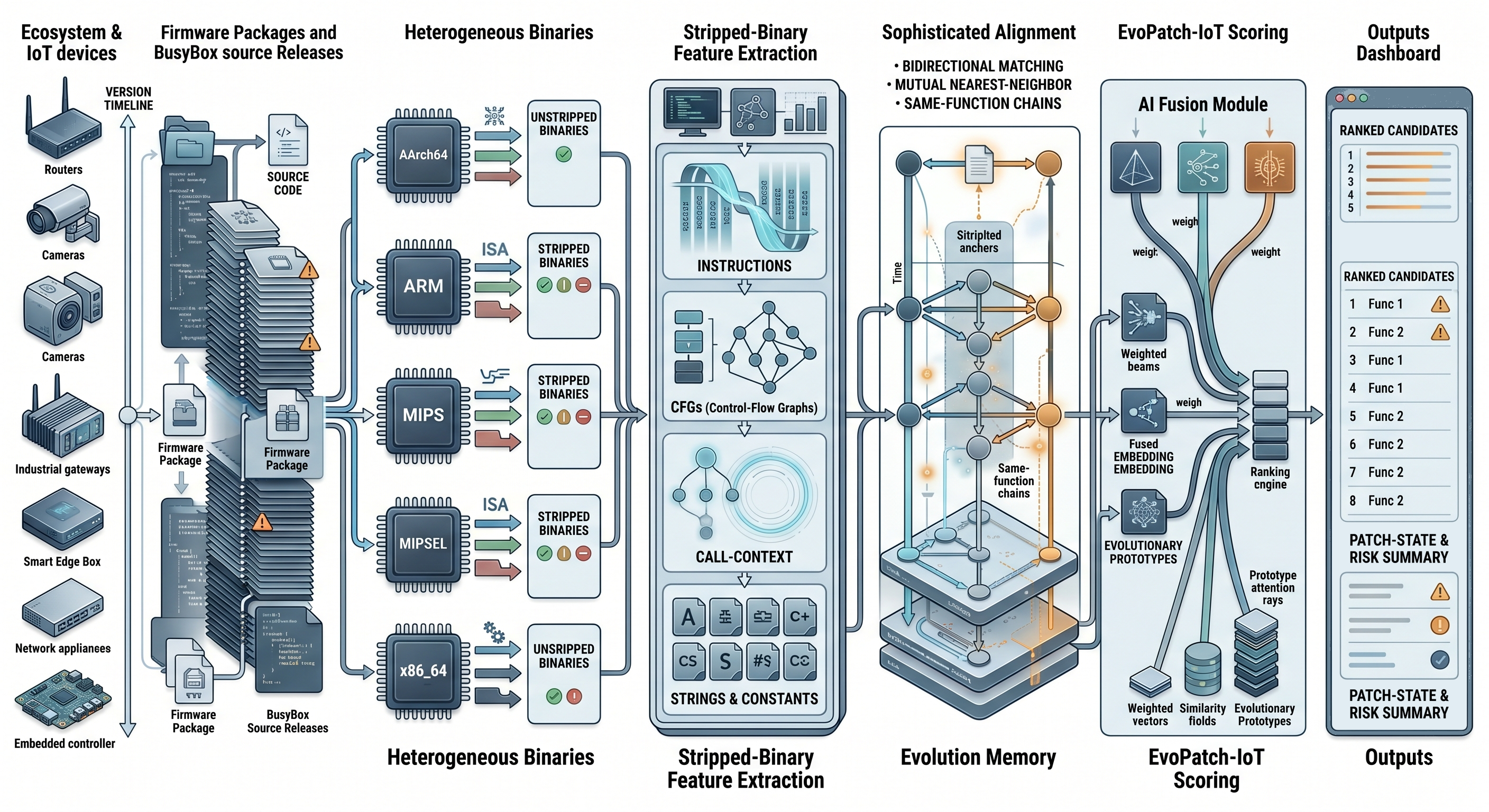}
\caption{Overview of EvoPatch-IoT. The pipeline starts from heterogeneous IoT firmware ecosystems and long-term BusyBox source releases, compiles paired stripped and unstripped binaries across multiple ISAs, recovers anonymous binary evidence through a symbol-free analysis engine, establishes bidirectional stripped-to-labeled anchors across versions, and finally fuses geometric shape priors, anonymous multi-view embeddings, and historical evolution prototypes to rank vulnerable function candidates and support patch-state evidence reporting.}
\label{fig:overview}
\end{figure*}

\begin{figure*}[!t]
\centering
\includegraphics[width=0.98\textwidth]{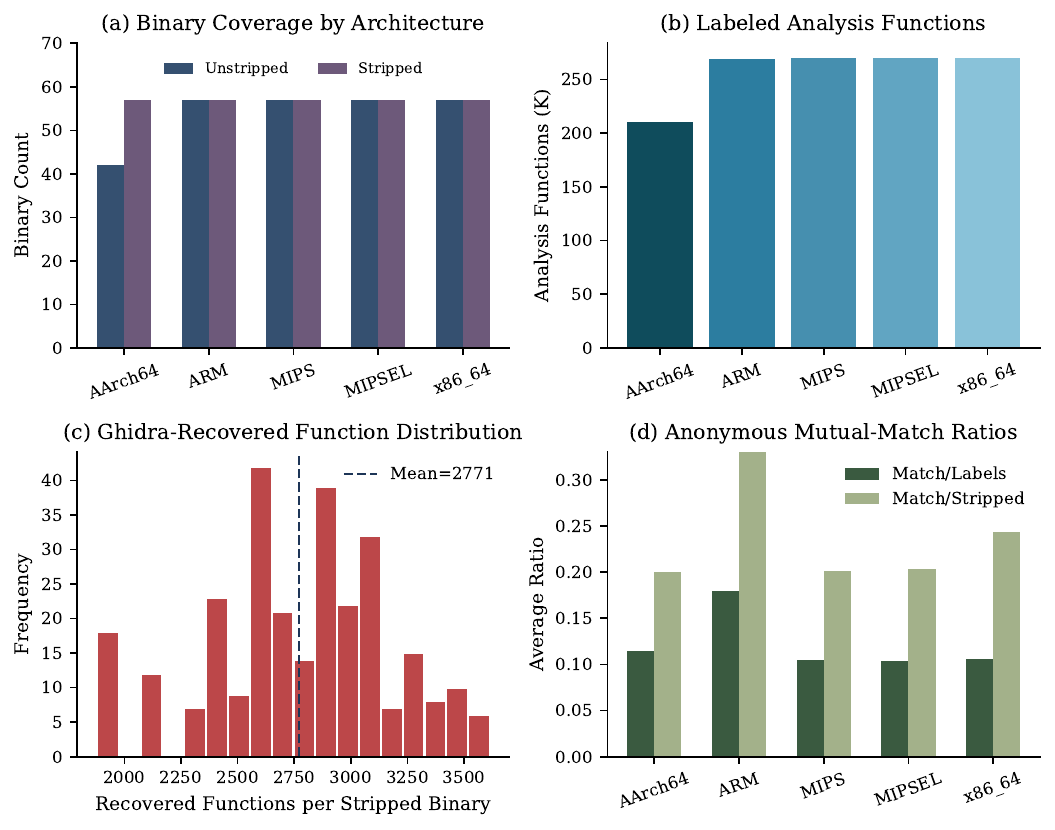}
\caption{Benchmark overview. Our dataset jointly exposes architecture coverage, large-scale analysis-function labels, stripped-function recovery statistics, and anonymous mutual-match ratios required by the alignment layer.}
\label{fig:dataset}
\end{figure*}

\subsection{Tasks}
We formalize three progressively richer tasks.

\textbf{Task 1: Cross-architecture function retrieval.} Given a stripped query function $q$ from architecture $a_i$ and a candidate set $\mathcal{C}_{a_j}$ from a different architecture $a_j$ under the same BusyBox version, rank homologous functions as high as possible.

\textbf{Task 2: Evolution-aware vulnerability retrieval.} Given a historical vulnerable or reference function and a stripped candidate pool from another version or architecture, rank functions that preserve the same upstream semantics while remaining robust to version drift.

\textbf{Task 3: Binary-level patch-state profiling.} Given a target binary and a CVE-related reference pattern, infer whether the target behaves more like the vulnerable or patched lineage, and expose a compact set of evidence functions.

\subsection{Evaluation Metrics}
For retrieval tasks we report Hit@1, Hit@5, Hit@10, MRR@10, and Mean Inspected@10. Let $r(q)$ denote the rank of the first correct candidate for query $q$, clipped at 11 if no correct result appears in the top 10. Then
\begin{equation}
\mathrm{MRR@10} = \frac{1}{|\mathcal{Q}|}\sum_{q \in \mathcal{Q}} \frac{\mathbb{I}[r(q)\leq 10]}{r(q)}.
\end{equation}
Here $\mathcal{Q}$ is the query set, $|\mathcal{Q}|$ is the number of evaluated queries, $\mathbb{I}[\cdot]$ is the indicator function that equals $1$ when its condition is true and $0$ otherwise, and $r(q)$ is the first-hit rank of query $q$ after top-10 clipping.
We also use the inspection-reduction view that is natural for firmware auditing:
\begin{equation}
\mathrm{InspectionReduction} = 1 - \frac{\mathbb{E}[r(q)]}{\mathbb{E}[|\mathcal{C}|]}.
\end{equation}
In this expression, $\mathbb{E}[\cdot]$ denotes expectation over the evaluated queries, $r(q)$ is again the inspected rank of the first correct answer, $\mathcal{C}$ is the candidate pool for the corresponding retrieval task, and $|\mathcal{C}|$ is the candidate-pool size.
For patch-state profiling we report accuracy, precision, recall, and F1.

\section{EvoPatch-IoT Method}
\subsection{Overview}
EvoPatch-IoT is deliberately split into four layers: stripped-compatible feature extraction, anonymous alignment, multi-view representation construction, and evolution-aware retrieval. Figure~\ref{fig:overview} illustrates the complete architecture. The left side begins with a realistic IoT firmware ecosystem, where routers, cameras, industrial gateways, embedded controllers, and network appliances reuse BusyBox over long release timelines. From these artifacts, the benchmark branch organizes BusyBox source releases, firmware packages, and both unstripped and stripped binaries across AArch64, ARM, MIPS, MIPSEL, and x86\_64. This stage provides the longitudinal version coverage required for evolution-aware analysis while preserving the stripped-binary setting encountered in deployed firmware.

The middle of Fig.~\ref{fig:overview} corresponds to the executable analysis pipeline. The binary analysis engine recovers anonymous functions without relying on symbols and extracts normalized instruction tokens, control-flow graphs, call-context features, and string/constant categories. These views are converted into shape, graph, token, and context representations. The sophisticated alignment zone then performs bidirectional nearest-neighbor matching between stripped functions and labeled historical functions, producing high-confidence anchors and same-function chains across versions. The right side implements the final fusion and ranking stage: geometric shape similarity provides the local backbone, anonymous multi-view embedding similarity resolves semantic ambiguity, and historical prototype similarity injects version-evolution memory. The weighted ranking engine outputs ranked vulnerable-function candidates, patch-state decisions, supporting evidence functions, firmware risk summaries, and CVE-oriented change regions for analyst inspection.

\subsection{Stripped-Compatible Anonymous Features}
For each recovered stripped function $f$, we build a geometric shape descriptor from its location inside the current binary:
\begin{equation}
\bm{s}(f) =
\big[\ell(f),\, \eta_{\mathrm{addr}}(f),\, \eta_{\mathrm{rank}}(f),\, \bar{\ell}_{\mathcal{N}_2(f)},\, \Delta \ell(f)\big]^{\top},
\label{eq:shape}
\end{equation}
where $\ell(f)=\log(1+\mathrm{size}(f))$, $\eta_{\mathrm{addr}}(f)=\frac{\mathrm{addr}(f)-a_{\min}}{a_{\max}-a_{\min}}$, $\eta_{\mathrm{rank}}(f)=\frac{\mathrm{rank}(f)}{N-1}$, and $\Delta \ell(f)=\ell(f)-\bar{\ell}_{\mathcal{N}_2(f)}$. Here $\mathrm{size}(f)$ is the recovered function size in bytes, $\mathrm{addr}(f)$ is the start address of $f$, $a_{\min}$ and $a_{\max}$ are the minimum and maximum recovered function addresses in the current binary, $\mathrm{rank}(f)$ is the address-sorted position of $f$, $N$ is the number of recovered functions in that binary, $\mathcal{N}_2(f)$ is the two-sided local neighborhood around $f$, and $\bar{\ell}_{\mathcal{N}_2(f)}$ is the mean log-size inside that neighborhood. This five-dimensional vector is crucial because stripped-to-unstripped correspondence is otherwise ill-posed in the absence of names.

The Ghidra exporter additionally generates anonymous instruction tokens, graph statistics, call contexts, string categories, and constant buckets. Let $\mathcal{T}(f)$ be the token multiset and $\mathcal{X}(f)$ be the context multiset. We convert both into TF-IDF weighted sparse vectors and then hash them into fixed-length dense embeddings:
\begin{equation}
\begin{aligned}
\bm{t}(f)&=H_{256}(\mathrm{TFIDF}(\mathcal{T}(f))), \\
\bm{c}(f)&=H_{64}(\mathrm{TFIDF}(\mathcal{X}(f))).
\end{aligned}
\end{equation}
Here $\mathcal{T}(f)$ is the multiset of anonymous instruction-like tokens extracted from function $f$, $\mathcal{X}(f)$ is the multiset of anonymous context events such as calls, strings, and constants, $\mathrm{TFIDF}(\cdot)$ converts a multiset into a sparse term-weight vector, $H_{256}(\cdot)$ and $H_{64}(\cdot)$ are fixed-dimensional hashing projections, and $\bm{t}(f)$ and $\bm{c}(f)$ are the resulting token and context embeddings.

Graph and structural statistics are summarized in two parts. First,
\begin{equation}
\bm{g}_{\mathrm{base}}(f)=
\big[\bm{\ell}(f)\Vert \bm{\rho}(f)\Vert \bar{i}_{bb}\big],
\end{equation}
where
\[
\bm{\ell}(f)=\big[\log(1+s_f), \log(1+i_f), \log(1+b_f), \log(1+e_f)\big]
\]
and
\[
\bm{\rho}(f)=\big[\rho_{\mathrm{call}}, \rho_{\mathrm{branch}}, \rho_{\mathrm{ret}}, \rho_{\mathrm{str}}, \rho_{\mathrm{const}}, \rho_{\mathrm{edge}}\big].
\]
We then concatenate these summary terms with histogram views:
\begin{equation}
\begin{aligned}
\bm{g}(f)=&\bm{g}_{\mathrm{base}}(f)\Vert \bm{h}_{op}(f) \\
&\Vert \bm{h}_{edge}(f).
\end{aligned}
\end{equation}
In these expressions, $\Vert$ denotes vector concatenation, $s_f$ is the byte size of function $f$, $i_f$ is its instruction count, $b_f$ is its number of basic blocks, $e_f$ is its number of CFG edges, $\rho_{\mathrm{call}}$, $\rho_{\mathrm{branch}}$, $\rho_{\mathrm{ret}}$, $\rho_{\mathrm{str}}$, $\rho_{\mathrm{const}}$, and $\rho_{\mathrm{edge}}$ are normalized densities of call sites, branches, returns, string references, constant references, and outgoing control-flow edges, respectively, $\bar{i}_{bb}$ is the mean number of instructions per basic block, and $\bm{h}_{op}(f)$ and $\bm{h}_{edge}(f)$ are the operation-class and edge-type histograms.

\subsection{Bidirectional Anonymous Alignment}
The unstripped branch provides normalized function identifiers used only for supervision and evaluation. To connect stripped functions with these labels without leakage, EvoPatch-IoT performs a bidirectional mutual nearest-neighbor match inside each $(v,a)$ bucket. The shape distance is
\begin{equation}
D_s(f_i,f_j) = \sum_{k=1}^{5}\left(\frac{s_k(f_i)-s_k(f_j)}{\alpha_k}\right)^2,
\label{eq:shape-distance}
\end{equation}
where $f_i$ and $f_j$ are two functions being compared, $s_k(f)$ is the $k$-th component of the shape vector $\bm{s}(f)$ in Eq.~(\ref{eq:shape}), the index $k$ runs over the five shape dimensions, and $\bm{\alpha}=[1.0,0.20,0.20,1.0,1.0]$ is the positive scale vector used to normalize the contribution of each dimension in the implementation.

Algorithm~\ref{alg:align} builds high-confidence anonymous supervision anchors between stripped and unstripped binaries that share the same version-architecture bucket. Line~1 declares the labeled function set $\mathcal{L}_{v,a}$, the stripped function set $\mathcal{S}_{v,a}$, the local search-window width $w$, and the acceptance threshold $\delta$. Lines~2--5 scan each labeled function $\ell$, retrieve a nearby stripped candidate window $\mathcal{C}_{\ell}$ centered on the primary geometric coordinate $s_1(\ell)$, and record the nearest stripped neighbor $b_{\mathcal{S}}(\ell)$. Lines~6--9 repeat the same operation in the reverse direction, now starting from each stripped function $s$ and storing its nearest labeled neighbor $b_{\mathcal{L}}(s)$. Line~10 initializes the final match set $\mathcal{M}$. Lines~11--15 enforce the core mutual-consistency rule: a pair is accepted only if the forward and reverse nearest neighbors agree and the corresponding shape distance does not exceed $\delta$. Line~16 returns the accepted anonymous anchors, which are then used for supervision, prototype construction, and evaluation bookkeeping.

This process yields 155,845 mutual matches across 270 paired binaries, averaging 577.2 anchors per binary. Figure~\ref{fig:dataset}(d) shows that the match ratio is architecture dependent, which further justifies retaining a geometry-aware retrieval backbone in the final scoring rule.

\subsection{Architecture-Normalized Multi-View Fusion}
Once anonymous anchors are available, we normalize graph and shape vectors with statistics from the training split. Let $\mu_a^{(g)}, \sigma_a^{(g)}$ be per-architecture moments for graph features, and similarly $\mu_a^{(s)}, \sigma_a^{(s)}$ for shape features:
\begin{equation}
\begin{aligned}
\tilde{\bm{g}}(f) &= \frac{\bm{g}(f)-\mu^{(g)}_{a(f)}}{\sigma^{(g)}_{a(f)}+\epsilon}, \\
\tilde{\bm{s}}(f) &= \frac{\bm{s}(f)-\mu^{(s)}_{a(f)}}{\sigma^{(s)}_{a(f)}+\epsilon}.
\end{aligned}
\end{equation}
Here $a(f)$ is the architecture label of function $f$, $\mu^{(g)}_{a(f)}$ and $\sigma^{(g)}_{a(f)}$ are the training-split mean and standard-deviation vectors for graph features under architecture $a(f)$, $\mu^{(s)}_{a(f)}$ and $\sigma^{(s)}_{a(f)}$ are the corresponding moments for shape features, $\tilde{\bm{g}}(f)$ is the normalized graph vector, $\tilde{\bm{s}}(f)$ is the normalized shape vector, and $\epsilon$ is a small numerical-stability constant.
The final fused representation is the concatenation
\begin{equation}
\bm{z}(f) = \bm{t}(f) \,\Vert\, \tilde{\bm{g}}(f) \,\Vert\, \bm{c}(f) \,\Vert\, \tilde{\bm{s}}(f).
\label{eq:fusion}
\end{equation}
In Eq.~(\ref{eq:fusion}), $\Vert$ again denotes concatenation, $\bm{t}(f)$ is the token embedding, $\tilde{\bm{g}}(f)$ is the normalized graph vector, $\bm{c}(f)$ is the context embedding, $\tilde{\bm{s}}(f)$ is the normalized shape vector, and $\bm{z}(f)\in\mathbb{R}^{361}$ is the final fused representation. In the current implementation, $\bm{z}(f)$ is a deterministic 361-dimensional multi-view representation rather than a black-box end-to-end embedding. This design keeps the pipeline reproducible and transparent while still allowing strong retrieval performance.

\begin{algorithm}[t]
\caption{Bidirectional Anonymous Alignment}
\label{alg:align}
\begin{algorithmic}[1]
\REQUIRE labeled functions $\mathcal{L}_{v,a}$, stripped functions $\mathcal{S}_{v,a}$, search window $w$, distance threshold $\delta$
\FOR{each $\ell \in \mathcal{L}_{v,a}$}
    \STATE $\mathcal{C}_{\ell} \leftarrow \mathrm{TopWindow}(\mathcal{S}_{v,a}, s_1(\ell), w)$
    \STATE $b_{\mathcal{S}}(\ell) \leftarrow \arg\min_{s \in \mathcal{C}_{\ell}} D_s(\ell,s)$
\ENDFOR
\FOR{each $s \in \mathcal{S}_{v,a}$}
    \STATE $\mathcal{C}_{s} \leftarrow \mathrm{TopWindow}(\mathcal{L}_{v,a}, s_1(s), w)$
    \STATE $b_{\mathcal{L}}(s) \leftarrow \arg\min_{\ell \in \mathcal{C}_{s}} D_s(s,\ell)$
\ENDFOR
\STATE $\mathcal{M} \leftarrow \emptyset$
\FOR{each $s \in \mathcal{S}_{v,a}$}
    \IF{$D_s(s,b_{\mathcal{L}}(s)) \leq \delta$ and $b_{\mathcal{S}}(b_{\mathcal{L}}(s)) = s$}
        \STATE add $(s,b_{\mathcal{L}}(s))$ to $\mathcal{M}$
    \ENDIF
\ENDFOR
\RETURN $\mathcal{M}$
\end{algorithmic}
\end{algorithm}

\subsection{Historical Evolution Prototypes}
For each normalized function identity $u$, we aggregate training functions from older versions into a historical prototype:
\begin{equation}
\bm{p}_u = \frac{1}{|\mathcal{T}_u|}\sum_{f \in \mathcal{T}_u} \bm{z}(f),
\label{eq:prototype}
\end{equation}
where $u$ is a normalized function identity, $\mathcal{T}_u$ is the set of matched training functions with identity $u$, $|\mathcal{T}_u|$ is the number of such functions, $\bm{z}(f)$ is the fused vector of function $f$, and $\bm{p}_u$ is the resulting historical prototype. These prototypes encode long-term version behavior and act as a stabilizer when the query and candidate are separated by substantial version drift.

\begin{algorithm}[t]
\caption{Evolution-Aware Retrieval Scoring}
\label{alg:score}
\begin{algorithmic}[1]
\REQUIRE query $q$, candidate pool $\mathcal{C}$, prototype bank $\mathcal{P}$
\STATE build $\bm{s}(q)$ and $\bm{z}(q)$
\FOR{each $c \in \mathcal{C}$}
    \STATE compute $\bm{s}(c)$ and $\bm{z}(c)$
    \STATE $r_s(q,c) \leftarrow -D_s(q,c)$
    \STATE $r_f(q,c) \leftarrow \cos(\bm{z}(q), \bm{z}(c))$
    \STATE $r_p(q,c) \leftarrow \cos(\bm{z}(c), \bm{p}_{u(q)})$ if prototype exists, else $0$
    \STATE $R(q,c) \leftarrow 0.70\,r_s(q,c) + 0.10\,r_f(q,c) + 0.20\,r_p(q,c)$
\ENDFOR
\RETURN candidates ranked by $R(q,c)$
\end{algorithmic}
\end{algorithm}

Algorithm~\ref{alg:score} implements the final evolution-aware ranking rule used in deployment. Line~1 specifies the stripped query $q$, the candidate pool $\mathcal{C}$ from the target binary, and the historical prototype bank $\mathcal{P}$. Line~2 computes the query-side shape descriptor and fused vector only once. Lines~3--9 iterate over every candidate $c \in \mathcal{C}$. Line~4 constructs the candidate-side shape and fused representations. Line~5 evaluates the geometry score $r_s(q,c)$ through the negative shape distance. Line~6 computes the fused multi-view similarity $r_f(q,c)$ via cosine similarity in the 361-dimensional space. Line~7 adds the historical evolution term $r_p(q,c)$ by comparing the candidate to the prototype associated with the query identity whenever such a prototype exists. Line~8 combines the three partial scores into the final retrieval score $R(q,c)$ using fixed weights. Line~10 returns the candidates sorted by this score, which directly produces the ranked list inspected by analysts.

Equation~(\ref{eq:prototype}) and Algorithm~\ref{alg:score} correspond to the deployed EvoPatch-IoT scorer:
\begin{equation}
\begin{aligned}
R(q,c)=&\lambda_s\left[-D_s(q,c)\right] + \lambda_f \cos(\bm{z}(q),\bm{z}(c)) \\
&+ \lambda_p \cos(\bm{z}(c),\bm{p}_{u(q)}),
\end{aligned}
\label{eq:score}
\end{equation}
In Eq.~(\ref{eq:score}), $q$ is the query function, $c$ is a candidate function from the target pool, $D_s(q,c)$ is the shape distance from Eq.~(\ref{eq:shape-distance}), $\cos(\cdot,\cdot)$ is cosine similarity, $\bm{z}(q)$ and $\bm{z}(c)$ are the fused vectors of the query and candidate, $u(q)$ is the normalized identity attached to the query during supervision, $\bm{p}_{u(q)}$ is the corresponding historical prototype, and $(\lambda_s,\lambda_f,\lambda_p)=(0.70,0.10,0.20)$ are the geometry, fusion, and prototype weights, respectively. The weighting reflects an important empirical observation: under anonymous alignment, local function geometry is the most reliable retrieval backbone, while multi-view fusion and evolution prototypes are best used to rerank ambiguous neighborhoods.

\subsection{Binary-Level Patch-State Proxy}
To approximate patch-state profiling at the binary level, we additionally build a lightweight proxy classifier from whole-binary statistics:
\begin{equation}
\bm{m}(b) = \left[
\begin{array}{c}
\mathrm{size}_{MB},\, n_{\mathrm{sym}},\, n_{\mathrm{ana}},\, \mu_{\mathrm{size}},\, \mathrm{med}_{\mathrm{size}}, \\[2pt]
p90_{\mathrm{size}},\, p99_{\mathrm{size}},\, n_{\mathrm{sec}},\, n_{\mathrm{debug}}
\end{array}
\right].
\end{equation}
Here $b$ denotes a whole BusyBox binary, $\mathrm{size}_{MB}$ is its file size in megabytes, $n_{\mathrm{sym}}$ is the number of available function symbols in the metadata branch, $n_{\mathrm{ana}}$ is the number of Ghidra-recovered analysis functions, $\mu_{\mathrm{size}}$ is the mean recovered function size, $\mathrm{med}_{\mathrm{size}}$ is the median recovered function size, $p90_{\mathrm{size}}$ and $p99_{\mathrm{size}}$ are the 90th and 99th percentiles of recovered function sizes, $n_{\mathrm{sec}}$ is the number of binary sections, and $n_{\mathrm{debug}}$ is the number of debug-related sections or debug-bearing records retained by the loader.
For a specific CVE label $y \in \{0,1\}$, the proxy predicts the closest standardized class centroid:
\begin{equation}
\hat{y}(b)=\arg\min_{y \in \{0,1\}} \left\| \frac{\bm{m}(b)-\mu_y}{\sigma_y+\epsilon} \right\|_2.
\end{equation}
In this classifier, $\hat{y}(b)$ is the predicted patch-state label of binary $b$, the two values of $y$ correspond to the vulnerable and patched classes, $\mu_y$ and $\sigma_y$ are the class-specific mean and standard-deviation vectors estimated from training binaries, $\epsilon$ is a small stabilizer, and $\|\cdot\|_2$ is the Euclidean norm.
We use this proxy only as an auxiliary binary-level study; the core contribution of the paper remains the function-level evolution-aware retrieval layer.

\subsection{Complexity and Deployment Profile}
The current pipeline is intentionally lightweight. Anonymous alignment uses a bounded search window instead of full quadratic pairwise matching. If $w$ is the shape-window width, then the matching cost per paired binary is
\begin{equation}
T_{\mathrm{align}} = O((|\mathcal{L}|+|\mathcal{S}|)\cdot w),
\end{equation}
where $T_{\mathrm{align}}$ is the alignment cost for one paired binary, $|\mathcal{L}|$ is the number of labeled functions, $|\mathcal{S}|$ is the number of stripped functions, and $w$ is the bounded local search-window width. This bound holds because each labeled or stripped function only compares against a local neighborhood around the primary shape coordinate. After alignment, ranking reduces to dense similarity scoring:
\begin{equation}
T_{\mathrm{rank}} = O(|\mathcal{Q}|\cdot|\mathcal{C}|\cdot d),
\end{equation}
where $T_{\mathrm{rank}}$ is the ranking cost, $|\mathcal{Q}|$ is the number of queries, $|\mathcal{C}|$ is the candidate-pool size, and $d=361$ is the fused feature dimension. In our implementation, $w=96$, the weighted candidate pool size is 609.41 functions, and EvoPatch-IoT inspects only 6.20 functions on average before hitting a correct top-10 candidate. This makes the framework practical for firmware triage pipelines that must scan large device fleets quickly.

\section{Experimental Setup}
\subsection{Compared Methods}
All comparison methods operate under the same stripped-compatible protocol and consume the same anonymous feature source. We evaluate 13 method families: SizeStat, ShapeStat, CLAP, ISSTA 2024, GTrans, BAR 2024, AMMF, Cybersecurity 2025, Binary2vec, Array 2025, VEXIR2Vec 2023/2025-style, Ex2Vec 2025, and our EvoPatch-IoT. These are unified reproductions of representative modeling biases rather than paper-for-paper engineering replicas, which is intentional because our goal is a fair, artifact-consistent comparison under a shared firmware-oriented protocol.

Two baselines deserve explicit clarification. \emph{SizeStat} and \emph{ShapeStat} are not claimed as new methods of this paper; they are simple in-house heuristic baselines implemented inside our current stripped-compatible evaluation framework so that all methods can be compared under exactly the same artifact pipeline. For a query $q$ and candidate $c$, SizeStat uses only the recovered function size and ranks candidates by
\begin{equation}
r_{\mathrm{size}}(q,c)=-\left|\log(1+\mathrm{size}(q))-\log(1+\mathrm{size}(c))\right|.
\end{equation}
In this score, $q$ is the query function, $c$ is the candidate function, $\mathrm{size}(\cdot)$ is the recovered function size in bytes, the logarithm suppresses extreme size variation, and a larger value of $r_{\mathrm{size}}(q,c)$, that is, a smaller size gap, indicates a better match. 
ShapeStat extends this lower bound to anonymous local geometry by directly reusing the five-dimensional descriptor of Eq.~(\ref{eq:shape}) and the geometric distance in Eq.~(\ref{eq:shape-distance}):
\begin{equation}
r_{\mathrm{shape}}(q,c)=-D_s(q,c).
\end{equation}
Here $D_s(q,c)$ is the shape distance between $q$ and $c$, and again a larger score means that the two functions are geometrically more similar. 
We include these two controls for two reasons. First, stripped BusyBox auditing often begins with extremely weak evidence, so a realistic benchmark should reveal how much performance can already be obtained from size-only and geometry-only cues before semantic modeling is introduced. Second, these baselines isolate the exact gain contributed by fusion and historical evolution priors. In particular, ShapeStat tells us whether EvoPatch-IoT is truly benefiting from evolution-aware reranking rather than merely inheriting the anonymous alignment geometry. Only \emph{EvoPatch-IoT} is the proposed method of this paper.

\subsection{Protocol}
The all-version evaluation is driven by the run \path{codes_all57_eval_full_plus_vex_ex_20260419}. It covers 57 versions, 1,020 directed architecture pairs, and 128,084 evaluated query functions. The Ghidra source run fully covers all five architectures for all 57 versions. Functions with fewer than 16 bytes or four instructions are filtered out. Matching uses a search window of 96 candidate neighbors and a maximum shape distance of 0.20.

\subsection{Implementation Details and Cost}
The artifact stack is reproducible from the local project workspace. Symbol extraction uses \texttt{readelf}, stripped feature extraction uses headless Ghidra, and the all-version comparison stage uses a deterministic Python scorer without downloading pretrained checkpoints. Across the 285 stripped binaries, the full Ghidra export takes 5.04 hours in total. Mean per-binary extraction time is 43.49 s for AArch64, 58.63 s for ARM, 82.19 s for MIPS, 84.31 s for MIPSEL, and 49.96 s for x86\_64. These numbers are small enough for periodic firmware monitoring and also reveal an operational reality: the architectures that are harder to recover structurally often also cost more to process.

\subsection{Research Questions}
We organize the evaluation around six research questions.
\begin{itemize}
\item \textbf{RQ1:} How does EvoPatch-IoT compare with stripped-compatible baselines on all-version cross-architecture retrieval?
\item \textbf{RQ2:} Does the historical prototype improve robustness across long BusyBox version ranges?
\item \textbf{RQ3:} Which architecture pairs remain difficult, and where does the evolution-aware score help most?
\item \textbf{RQ4:} Can version-evolution signals transfer across architectures?
\item \textbf{RQ5:} Does the framework provide useful patch-state evidence on a real BusyBox vulnerability case?
\item \textbf{RQ6:} What deployment-efficiency patterns and failure modes emerge across architectures?
\end{itemize}

\section{Experimental Results}
\subsection{Alignment Quality Before Retrieval}
Before considering retrieval accuracy, it is important to inspect the quality of the anonymous stripped-to-unstripped bridge. Table~\ref{tab:alignment} summarizes this stage. ARM yields the highest average match ratio against labels (0.1792) and stripped functions (0.3944), while AArch64 attains the smallest median match distance (0.0358), suggesting that once a paired sample exists, its geometric anchors are especially stable. MIPS and MIPSEL recover fewer matches than ARM but remain highly consistent, which helps explain their strong pairwise retrieval behavior.

These observations also clarify why EvoPatch-IoT retains a geometry-aware backbone. Without a trustworthy anonymous bridge, historical prototypes and fused similarity cannot be exploited safely. EvoPatch-IoT is therefore best viewed as an \emph{alignment-aware} retrieval framework rather than a generic embedding model transplanted onto firmware binaries.

\begin{table}[t]
\caption{Anonymous Alignment Quality and Extraction Cost by Architecture}
\label{tab:alignment}
\centering
\footnotesize
\resizebox{\columnwidth}{!}{%
\begin{tabular}{lrrrrr}
\toprule
Arch & Matches & Match/Lbl & Match/Strp & Med.Dist & Time(s) \\
\midrule
AArch64 & 23,703 & 0.1149 & 0.2005 & 0.0358 & 43.49 \\
ARM & 48,229 & 0.1792 & 0.3944 & 0.0988 & 58.63 \\
MIPS & 28,419 & 0.1043 & 0.2012 & 0.0847 & 82.19 \\
MIPSEL & 28,443 & 0.1042 & 0.2031 & 0.0862 & 84.31 \\
x86\_64 & 27,051 & 0.1054 & 0.2430 & 0.0509 & 49.96 \\
\bottomrule
\end{tabular}
}
\end{table}

\subsection{RQ1: Overall Cross-Architecture Retrieval}
\begin{figure*}[!t]
\centering
\includegraphics[width=0.98\textwidth]{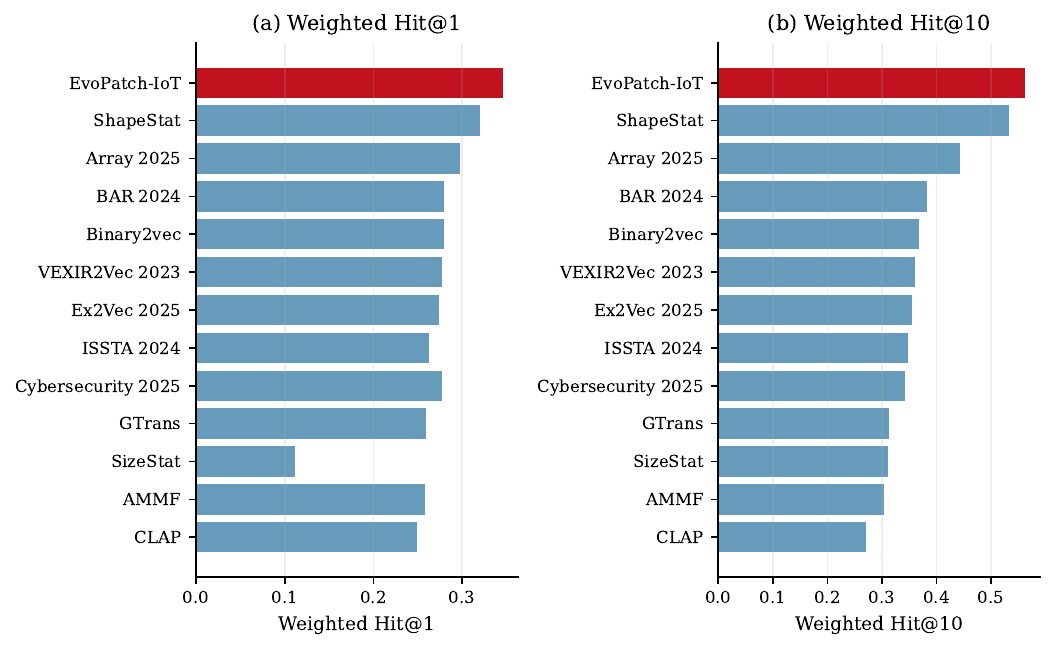}
\caption{Overall method ranking under the unified stripped-compatible protocol. EvoPatch-IoT leads both Hit@1 and Hit@10 across the 57-version evaluation.}
\label{fig:overall}
\end{figure*}

Table~\ref{tab:overall} and Figure~\ref{fig:overall} summarize the main comparison. EvoPatch-IoT achieves the best weighted Hit@1 (34.56\%), Hit@5 (48.70\%), Hit@10 (56.24\%), and MRR@10 (40.57\%). The strongest competing baseline is Array 2025 with 29.78\% Hit@1 and 44.33\% Hit@10, which means our method improves Hit@1 by 16.04\% and Hit@10 by 26.85\% relative. Against the strong non-neural ShapeStat baseline, the gains are still positive at 8.00\% for Hit@1 and 5.50\% for Hit@10.

The result is especially meaningful because ShapeStat is unusually strong in our setting: anonymous alignment is itself geometry-driven, and therefore size/order cues remain powerful. EvoPatch-IoT does not ignore this fact. Instead, it uses shape as the backbone and lets multi-view fusion plus evolution prototypes resolve ambiguity among near neighbors. This is exactly the design encoded in Eq.~(\ref{eq:score}).

The practical value is also clear. The weighted candidate pool size in our evaluation is 609.41 functions, whereas the weighted Mean Inspected@10 of EvoPatch-IoT is only 6.20. This corresponds to an inspection-reduction rate of 98.98\%, meaning auditors can reduce the candidate space by nearly two orders of magnitude before starting manual reverse engineering.

\begin{table*}[!t]
\caption{Overall Cross-Architecture Retrieval Results on 57 BusyBox Versions}
\label{tab:overall}
\centering
\scriptsize
\resizebox{\textwidth}{!}{%
\begin{tabular}{lccccccc}
\toprule
Method & Pairs & Queries & Weighted Hit@1 & Weighted Hit@5 & Weighted Hit@10 & Weighted MRR@10 & Mean Inspected@10 \\
\midrule
SizeStat & 1020 & 128,084 & 0.1122 & 0.2551 & 0.3106 & 0.1720 & 8.5597 \\
ShapeStat & 1020 & 128,084 & 0.3200 & 0.4532 & 0.5331 & 0.3771 & 6.5128 \\
CLAP~\cite{clap2024} & 1020 & 128,084 & 0.2493 & 0.2607 & 0.2701 & 0.2547 & 8.3887 \\
ISSTA 2024~\cite{clap2024,codenotnl2024} & 1020 & 128,084 & 0.2627 & 0.3043 & 0.3485 & 0.2825 & 7.9224 \\
GTrans~\cite{jtrans2022,rcfg2vec2024} & 1020 & 128,084 & 0.2589 & 0.2895 & 0.3133 & 0.2731 & 8.0938 \\
BAR 2024~\cite{strtune2024} & 1020 & 128,084 & 0.2791 & 0.3340 & 0.3822 & 0.3040 & 7.6405 \\
AMMF~\cite{palmtree2021,rcfg2vec2024} & 1020 & 128,084 & 0.2583 & 0.2850 & 0.3040 & 0.2702 & 8.1511 \\
Cybersecurity 2025~\cite{binquery2025} & 1020 & 128,084 & 0.2768 & 0.3128 & 0.3428 & 0.2934 & 7.8592 \\
Binary2vec~\cite{binary2vec2025} & 1020 & 128,084 & 0.2799 & 0.3294 & 0.3674 & 0.3021 & 7.7002 \\
Array 2025~\cite{binary2vec2025} & 1020 & 128,084 & 0.2978 & 0.3809 & 0.4433 & 0.3352 & 7.1848 \\
VEXIR2Vec 2023~\cite{vexir2vec2025} & 1020 & 128,084 & 0.2772 & 0.3238 & 0.3610 & 0.2983 & 7.7520 \\
Ex2Vec 2025~\cite{ex2vec2025} & 1020 & 128,084 & 0.2739 & 0.3188 & 0.3557 & 0.2943 & 7.7963 \\
\textbf{EvoPatch-IoT} & \textbf{1020} & \textbf{128,084} & \textbf{0.3456} & \textbf{0.4870} & \textbf{0.5624} & \textbf{0.4057} & \textbf{6.1957} \\
\bottomrule
\end{tabular}
}
\end{table*}

\subsection{RQ2: Version-Wise Robustness}
\begin{figure*}[!t]
\centering
\includegraphics[width=0.98\textwidth]{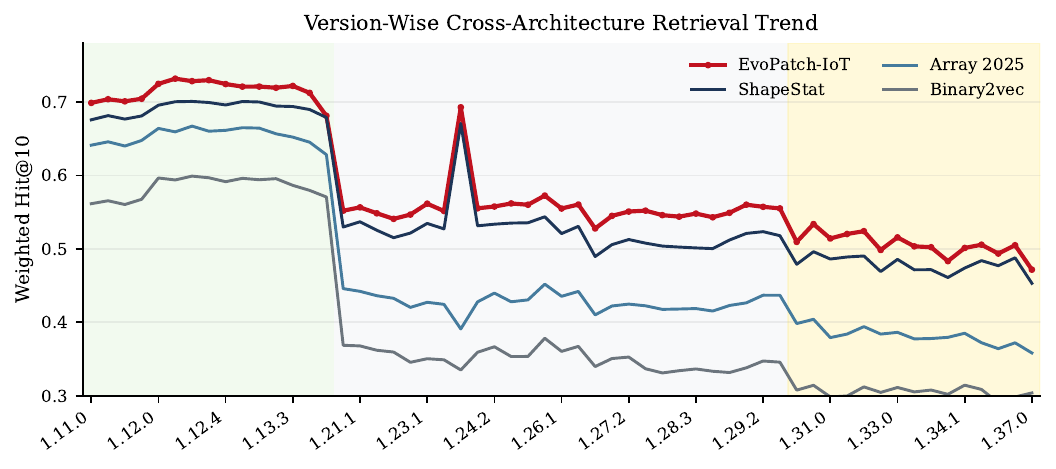}
\caption{Version-wise weighted Hit@10 on the 57-version benchmark. EvoPatch-IoT remains consistently above the strongest baselines from early to recent BusyBox releases.}
\label{fig:versiontrend}
\end{figure*}

EvoPatch-IoT is the best method on 56 of the 57 evaluated versions; ShapeStat is best only on BusyBox 1.14.0. Figure~\ref{fig:versiontrend} shows the version-wise Hit@10 curves. The absolute numbers decline from early legacy releases to the most recent releases because later BusyBox versions contain larger and structurally more diverse function spaces, but the margin of EvoPatch-IoT remains stable.

Table~\ref{tab:segments} groups the versions into legacy (1.11--1.14), middle (1.21--1.29), and recent (1.30--1.37) segments. EvoPatch-IoT keeps the best Hit@10 in all three segments, reaching 0.7150, 0.5575, and 0.5055, respectively. Compared with ShapeStat, the gains are +0.0240, +0.0319, and +0.0271 in absolute Hit@10. Compared with Array 2025, the gains are much larger, especially in the recent segment (+0.1244), which indicates that the historical prototype bank is helpful when version drift increases.

\subsection{Proxy Component Contribution Study}
The current artifact registry does not include a separate hand-crafted ablation run for every scorer component. However, because all methods are implemented under one stripped-compatible protocol, we can use the closest method families as a controlled \emph{proxy decomposition}. ShapeStat approximates a geometry-only backbone, Binary2vec approximates pure multi-view fusion, Array 2025 adds prototype-aware fusion without the explicit geometric backbone, and EvoPatch-IoT combines all three ingredients.

Table~\ref{tab:proxyablation} shows three clear trends. First, pure global fusion is not enough in the current stripped setting: Binary2vec underperforms ShapeStat because the anonymous bridge still depends heavily on local binary geometry. Second, prototype-aware fusion substantially recovers the lost performance, confirming that historical memory is genuinely helpful. Third, the best result comes from combining geometry, fusion, and prototypes in a single score, which is exactly the design of Eq.~(\ref{eq:score}).

\begin{table}[t]
\caption{Family-Level Proxy Study of EvoPatch-IoT Components}
\label{tab:proxyablation}
\centering
\scriptsize
\setlength{\tabcolsep}{2.4pt}
\resizebox{\columnwidth}{!}{%
\begin{tabular}{lccccc}
\toprule
Method & Geom. & Fusion & Proto. & Hit@10 & Mean Insp. \\
\midrule
ShapeStat & \checkmark &  &  & 0.5331 & 6.5128 \\
Binary2vec~\cite{binary2vec2025} &  & \checkmark &  & 0.3674 & 7.7002 \\
Array 2025~\cite{binary2vec2025} &  & \checkmark & \checkmark & 0.4433 & 7.1848 \\
\textbf{EvoPatch-IoT} & \checkmark & \checkmark & \checkmark & \textbf{0.5624} & \textbf{6.1957} \\
\bottomrule
\end{tabular}
}
\end{table}

\begin{table*}[!t]
\caption{Version-Segment Robustness of Representative Methods}
\label{tab:segments}
\centering
\small
\begin{tabular}{lcccccc}
\toprule
Method & Legacy H@1 & Legacy H@10 & Middle H@1 & Middle H@10 & Recent H@1 & Recent H@10 \\
\midrule
\textbf{EvoPatch-IoT} & \textbf{0.5280} & \textbf{0.7150} & \textbf{0.3315} & \textbf{0.5575} & \textbf{0.2919} & \textbf{0.5055} \\
ShapeStat & 0.5160 & 0.6910 & 0.3027 & 0.5256 & 0.2657 & 0.4784 \\
Array 2025~\cite{binary2vec2025} & 0.5102 & 0.6533 & 0.2763 & 0.4277 & 0.2440 & 0.3811 \\
Binary2vec~\cite{binary2vec2025} & 0.4890 & 0.5837 & 0.2583 & 0.3501 & 0.2274 & 0.3052 \\
\bottomrule
\end{tabular}
\end{table*}

\subsection{RQ3: Architecture-Pair Analysis}
\begin{figure*}[!t]
\centering
\includegraphics[width=0.98\textwidth]{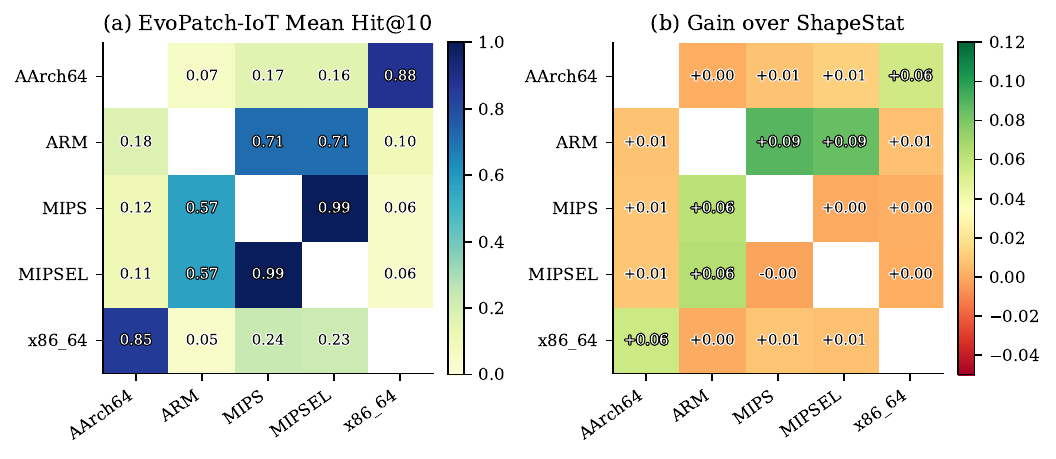}
\caption{Architecture-pair retrieval analysis. Left: mean Hit@10 of EvoPatch-IoT across all versions. Right: absolute Hit@10 gain over the strong ShapeStat baseline.}
\label{fig:heatmap}
\end{figure*}

Figure~\ref{fig:heatmap} reports pairwise Hit@10 averages for EvoPatch-IoT and its absolute gain over ShapeStat. Unsurprisingly, the easiest transfers are MIPS$\leftrightarrow$MIPSEL, where Hit@10 reaches about 0.99 in both directions because instruction-set semantics are nearly identical. AArch64$\rightarrow$x86\_64 and x86\_64$\rightarrow$AArch64 are also surprisingly strong at 0.8791 and 0.8541, respectively, indicating that multi-view normalization successfully bridges ISA differences when the anonymous alignment quality is good.

The hardest directions involve ARM and x86\_64, particularly x86\_64$\rightarrow$ARM, whose mean Hit@10 is 0.0494. These are precisely the settings where the raw geometric prior is least sufficient. EvoPatch-IoT nevertheless improves over ShapeStat on almost all off-diagonal pairs and yields its largest absolute gains on ARM$\rightarrow$MIPS, ARM$\rightarrow$MIPSEL, AArch64$\rightarrow$x86\_64, and x86\_64$\rightarrow$AArch64.

\subsection{RQ4: Version-Change Transfer}
To isolate evolution signals, we reuse the earlier version-change transfer experiment that tests whether function-size changes observed in one architecture predict changed functions in another architecture across adjacent BusyBox versions. Across 801 directed transitions, the mean ROC-AUC is 0.9887 and the mean average precision is 0.9152. The positive changed rate is only 1.05\%, meaning the task is highly imbalanced and therefore non-trivial.

Interestingly, the best transfer occurs between MIPSEL$\rightarrow$MIPS and MIPS$\rightarrow$MIPSEL, with mean average precision above 0.98, but even difficult pairs such as ARM$\rightarrow$x86\_64 remain above 0.86 average precision. This observation supports the core thesis of EvoPatch-IoT: version evolution is not random noise but a reusable supervisory signal that travels across architectures.

\subsection{RQ5: Patch-State Profiling and CVE-2021-42386 Case Study}
We next turn to an upstream BusyBox vulnerability, CVE-2021-42386, which affects the AWK subsystem and is fixed starting from BusyBox 1.34.0. The AWK-oriented patch probe identifies 25 changed function instances out of 30 inspected function-architecture rows between versions 1.33.1 and 1.34.0. The largest relative changes consistently appear in \texttt{awk\_printf} and \texttt{awk\_main}, while small but repeated changes also occur in \texttt{awk\_exit}, \texttt{awk\_split}, and \texttt{awk\_getline}. This cross-architecture regularity is exactly the kind of signal the historical prototype bank is expected to leverage.

\begin{figure*}[!t]
\centering
\includegraphics[width=0.98\textwidth]{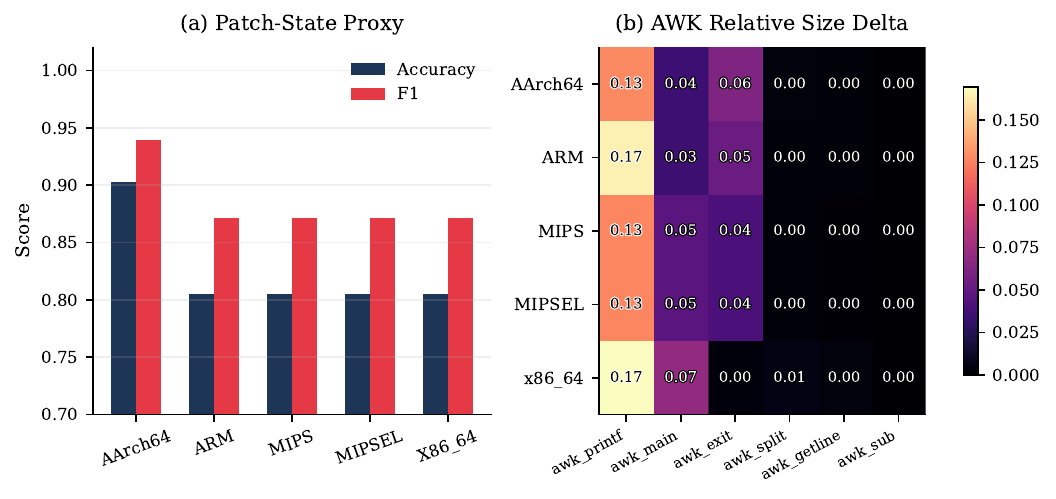}
\caption{Patch-state case study for CVE-2021-42386. Left: binary-level patch-state proxy across held-out architectures. Right: cross-architecture relative size deltas for representative AWK functions between BusyBox 1.33.1 and 1.34.0.}
\label{fig:patch}
\end{figure*}

Table~\ref{tab:patchproxy} and Figure~\ref{fig:patch} summarize the patch-state proxy. The binary-level nearest-centroid proxy achieves 82.44\% mean accuracy and 88.47\% mean F1 across held-out architectures, with perfect precision and 79.43\% mean recall. AArch64 is the strongest held-out architecture, reaching 90.24\% accuracy and 93.94\% F1. Although this proxy is simpler than the main retrieval engine, it demonstrates that even lightweight binary-level statistics already expose separable vulnerable-versus-patched behavior, and that function-level retrieval can provide concrete evidence for where the changes concentrate.

\begin{table}[t]
\caption{CVE-2021-42386 Patch-State Proxy Across Held-Out Architectures}
\label{tab:patchproxy}
\centering
\small
\begin{tabular}{lcccc}
\toprule
Held-Out Arch & Accuracy & Precision & Recall & F1 \\
\midrule
AArch64 & 0.9024 & 1.0000 & 0.8857 & 0.9394 \\
ARM & 0.8049 & 1.0000 & 0.7714 & 0.8710 \\
MIPS & 0.8049 & 1.0000 & 0.7714 & 0.8710 \\
MIPSEL & 0.8049 & 1.0000 & 0.7714 & 0.8710 \\
x86\_64 & 0.8049 & 1.0000 & 0.7714 & 0.8710 \\
\midrule
Mean & 0.8244 & 1.0000 & 0.7943 & 0.8847 \\
\bottomrule
\end{tabular}
\end{table}

\subsection{RQ6: Deployment Efficiency and Failure Modes}
\begin{figure*}[!t]
\centering
\includegraphics[width=0.98\textwidth]{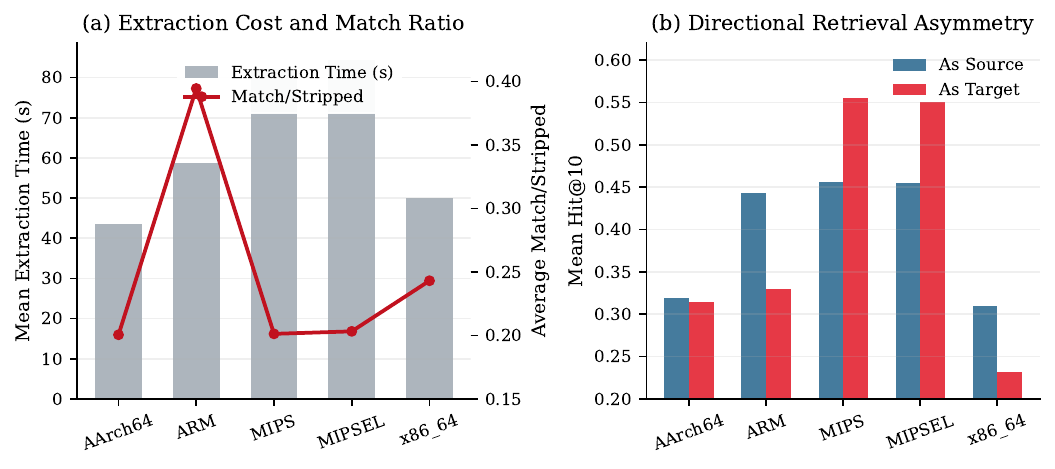}
\caption{Operational analysis of EvoPatch-IoT. Left: architecture-wise extraction time and anonymous match ratio. Right: average Hit@10 when each architecture acts as the source or the target side of cross-architecture retrieval.}
\label{fig:efficiency}
\end{figure*}

Figure~\ref{fig:efficiency} extends the main results from an operational viewpoint. The left panel combines mean Ghidra extraction time with anonymous match ratio. ARM attains the highest average match/stripped ratio (0.3944), but it is not the easiest target architecture. MIPS and MIPSEL require the longest extraction time (82.19\,s and 84.31\,s per binary on average), yet they are the easiest targets for function retrieval. Conversely, x86\_64 has only moderate extraction cost (49.96\,s) but is the hardest target overall. This means that raw decompilation cost and anonymous alignment density alone do not determine downstream retrieval difficulty; the target architecture's function-space organization matters at least as much.

The right panel reveals a strong directional asymmetry. When architectures act as sources, ARM, MIPS, and MIPSEL all stay around 0.44--0.46 mean Hit@10, whereas AArch64 and x86\_64 drop to 0.3192 and 0.3091. When they act as targets, MIPS and MIPSEL rise above 0.55 mean Hit@10, but x86\_64 falls to 0.2312. In other words, target-side candidate ambiguity is a larger bottleneck than query-side feature construction. This observation is consistent with our core design: Eq.~(\ref{eq:score}) gives the largest weight to geometry because the scorer must first localize a plausible target neighborhood before historical prototypes can help.

The difficult architecture directions also expose three concrete failure modes. First, x86\_64$\rightarrow$ARM reaches only 0.0494 mean Hit@10, showing that address rank and local-size geometry can drift sharply when compiler layout policies and calling conventions change simultaneously. Second, AArch64$\rightarrow$ARM remains low at 0.0656 even though AArch64 has the smallest median alignment distance, which indicates that a clean anonymous bridge is not sufficient if the target architecture contains many near-isomorphic helper functions. Third, MIPS/MIPSEL$\rightarrow$x86\_64 stays near 0.056--0.057, suggesting that prototype memory can still be confused by compact x86\_64 utility functions whose recovered CFG statistics become overly homogeneous after stripping. Even so, EvoPatch-IoT still outperforms ShapeStat on almost all off-diagonal directions, so the residual errors are best understood as boundary cases for future reranking and context-aware modeling rather than a failure of the evolution-aware formulation itself.

\section{Discussion}
The results lead to three practical observations.

First, stripped-binary geometry should not be dismissed as a weak heuristic in firmware analysis. In our environment it is the only reliable bridge between recovered stripped functions and unstripped labels, which is why ShapeStat is a strong baseline. EvoPatch-IoT succeeds not by replacing geometry, but by making geometry evolution-aware and multi-view consistent.

Second, the benchmark scale matters. The 57-version timeline and 155,845 mutual matches are large enough to reveal long-horizon trends that short three-version studies cannot. This is most visible in the segment analysis and the version-change transfer experiment.

Third, the current study still has limits. The binary-level patch-state classifier is a proxy rather than a fully learned function-to-binary aggregation network; BusyBox remains only one, albeit highly relevant, firmware component; and anonymous alignment quality directly affects upper-bound retrieval performance. These constraints should be viewed as realistic engineering boundaries rather than hidden assumptions. In future work, we plan to replace the handcrafted fusion rule with trainable ranking layers while preserving the same strict stripped-compatible evaluation policy.

\subsection{Implications for IoT Auditing Workflows}
From a deployment perspective, the main value of EvoPatch-IoT is not only higher top-$k$ accuracy, but also better evidence organization. The system produces a compact ranked list of stripped functions, historical prototypes for the queried identity, and architecture-pair statistics that indicate when an auditor should trust or doubt the returned neighborhood. This is useful for real firmware triage because reverse engineers rarely inspect an isolated function in a vacuum; they decide whether to continue based on how quickly the candidate list becomes semantically coherent.

Our results also suggest a practical scheduling strategy for fleet-scale monitoring. If an operator needs broad and cheap coverage, the longest architectural timeline should first be used to build prototype banks, while retrieval jobs targeting x86\_64 or ARM should be allocated a larger analyst budget because these directions remain more ambiguous. By contrast, MIPS$\leftrightarrow$MIPSEL transfers can serve as a high-confidence screening stage that rapidly flags likely reused vulnerable logic before more expensive manual confirmation.

\subsection{Threats to Validity}
Three validity threats deserve explicit discussion. First, the benchmark centers on BusyBox, which is highly relevant for IoT firmware but does not cover every component style. Second, our baselines are unified stripped-compatible reproductions rather than exact replicas of every original paper, so the comparison should be interpreted as a family-level study under one controlled protocol. Third, the patch-state experiment is still a proxy built from whole-binary statistics plus one representative CVE case. Even with these limitations, the benchmark is large, reproducible, and sufficiently challenging to support the main claim that evolution-aware retrieval materially improves function-level firmware auditing.

\section{Conclusion}
This paper presented EvoPatch-IoT, an evolution-aware cross-architecture retrieval framework for BusyBox-based IoT firmware. By combining stripped-compatible feature extraction, bidirectional anonymous alignment, architecture-normalized multi-view fusion, and historical function prototypes, EvoPatch-IoT delivers the best overall retrieval performance on a 57-version, five-architecture benchmark and remains robust under substantial version drift. The case-study results on CVE-2021-42386 further suggest that the same framework can support practical patch-state profiling. We believe the benchmark, algorithms, and quantitative findings provide a solid foundation for future IoT firmware auditing systems that require explainable function-level evidence from stripped binaries.

\section*{Acknowledgment}
This work is sponsored by the Guangdong Basic and Applied Basic Research Foundation (Grant Nos. 2025A1515010111 and 2025A1515010513) and the National Natural Science Foundation of China (Grant Nos. 72101059, 62406076, and 62506081).

\balance
\bibliographystyle{IEEEtran}
\bibliography{references}

\end{document}